\documentclass[fleqn,10pt]{wlscirep}
\usepackage[utf8]{inputenc}
\usepackage[T1]{fontenc}
\title{Speckled-speckle field as a resource for imaging techniques}

\author[1,+]{Silvia Cassina}
\author[1,2,+]{Gabriele Cenedese}
\author[1,3,*]{Alessia Allevi}
\author[3]{Maria Bondani}
\affil[1]{Department of Science and High Technology, University of Insubria, Via Valleggio 11, I-22100 Como, Italy}
\affil[2]{INFN- Section of Milan, Via Celoria 16, I-20133 Milano, Italy}
\affil[3]{Institute for Photonics and Nanotechnologies, IFN-CNR, Via Valleggio 11, I-22100 Como, Italy}

\affil[*]{alessia.allevi@uninsubria.it}

\affil[+]{these authors contributed equally to this work}

\begin{abstract}
Correlated states of light, both classical and quantum, can find useful applications in the implementation of several imaging techniques. Among the employed sources, pseudo-thermal states, generated by the passage of a laser beam through a diffuser, represent the standard choice. To produce light with a higher level of correlation, in this work we consider and characterize the speckled-speckle field obtained with two diffusers using both a numerical simulation and an experimental implementation. In order to discuss the potential usefulness of super-thermal light in imaging protocols, we analyze the behavior of some figures of merit, namely the contrast, the signal-to-noise ratio and the image resolution. The obtained results clarify the possible advantages offered by this kind of light, and at the same time better emphasize the reasons why it does not outperform pseudo-thermal light.\end{abstract}
\begin{document}

\flushbottom
\maketitle
%
%
\thispagestyle{empty}


\section*{Introduction}
The development of imaging techniques is currently acquiring a strong interest in view of their potential applications to medicine and biology \cite{samantaray,meda,ruoberchera}. One of the main issues in these fields is to prevent the sample from being damaged by the illuminating light, or altering chemical and biological photosensitive processes \cite{ivano,allevi22}. Therefore, two main solutions have been adopted over the years. One possibility is to operate in spectral regions in which the object is not affected by light absorptions \cite{lemos}, while the other one is to exploit correlated bipartite states of light, in which part of the field is strongly attenuated in order not to damage the sample, while the other arm is kept more intense and never interacts with the object \cite{ferri05}. The image is reconstructed by correlating the two parts of the field. Among the correlation-based techniques used so far we mention ghost imaging (GI), which exploits spatially-correlated beams \cite{dangelo,gatti}. One beam illuminates the object and then is sent to a detector without spatial resolution (bucket detector), while the other is addressed directly to a spatial-resolving detector, without interacting with the object. Neither of the two acquisitions separately contains information on the absorption profile of the object, which can be retrieved by means of correlations \cite{losero}. From the historical point of view, at first GI was experimentally implemented with entangled states of light \cite{pittman,abouraddy01,abouraddy04}, but later it has been demonstrated that the main resource required by the technique was the existence of correlations, so that also classically-correlated states of light could be used \cite{gatti04,gatti04a,bache04,bache04a,crosby,arimondo}. 
Many experiments have been performed with pseudo-thermal light, which is obtained by sending a laser beam through a moving diffuser, such as a rotating ground-glass disk \cite{arecchi}. At variance with entangled states of light, which exhibit perfect photon-number correlations but are quite fragile, classically-correlated states are more robust against loss, such as the ones introduced by a lower-than-1 detection efficiency \cite{ferri10}. Nevertheless, the use of nonclassical states results in a better image quality, as quantified by some figures of merit, such as the visibility and the signal-to-noise ratio (SNR) \cite{shapiro}. Some years ago Ferri et al. \cite{ferri10} proposed the so-called ``differential GI'' (DGI) exploiting pseudo-thermal light. The scheme addresses the problem of reconstructing small or faint objects, a case in which conventional GI fails because of the huge number of acquisitions required to reach a sufficiently high value of SNR. More recently, Losero et al. \cite{losero} applied the method to quantum states of light, demonstrating that also in this case DGI gives better results than GI, for any value of losses and light brightness.\\
In addition to pseudo-thermal light divided at a beam splitter (BS) and twin-beam states, which are both thermal in the photon-number statistics \cite{OE14,josaB14,pla22}, other kinds of correlated states have been suggested and tested in order to improve the values of the figures of merit of the image reconstruction \cite{dove,bromberg,bender,alves,li21}. In particular, it has been shown that light states, more correlated than the thermal ones, may lead to higher visibility, higher contrast or higher SNR \cite{iskhakov,ragy}. Among them, it is worth mentioning that the statistics of frequency-doubled thermal states is definitely ``super-thermal'', having it intensity fluctuations larger than those of a thermal field \cite{allevi15,allevi17}. Super-thermal statistics can be also obtained by using a sequence of diffusers \cite{odonnell,newman,yoshimura,zhou,liu21}, as some of us have experimentally demonstrated for two rotating ground-glass disks \cite{bianciardi} by measuring the photon-number statistics in the mesoscopic intensity regime by means of photon-number-resolving detectors \cite{OE17}.\\ 
In this work, we want to experimentally investigate the potential of super-thermal light for imaging applications by considering both GI and DGI schemes. After the preliminary characterization of the statistical properties, we prove that the exploitation of DGI instead of the standard GI yields better values of SNR, in analogy to what happens in the case of pseudo-thermal light, while there is no advantage in terms of contrast. 
We discuss advantages and limitations of our scheme by studying the dependence of the above-mentioned figures of merit on both the size of a binary object and the number of acquired images, for a fixed speckle dimension. In particular, we prove that super-thermal light allows the reconstruction of images with a high contrast for weakly absorbing objects, while guaranteeing good values of SNR.\\ 
Furthermore, we investigate whether the image resolution could benefit from the use of super-thermal light. The good quality of the results and the agreement between the theoretical model and the experimental outcomes, achieved with a numerical simulation and a real experiment, suggest a more practical exploitation of this kind of light, and encourage the use of more complex imaging protocols, in which the main features of super-thermal light can be fully exploited.

\section*{Results}
\subsection*{Theoretical description of speckled-speckle field}
A light field with super-thermal statistics can be generated by using a sequence of diffusers. It is well-known that a coherent light beam impinging on a diffuser produces a speckle field, composed of many coherence areas (speckles) that are the result of the constructive interference of the radiation diffused by the small random scattering centers contained in the illuminated area \cite{goodman}.\\ 
The statistics of light intensity corresponding to this speckle field is the thermal distribution\\ 
$p_{\rm th}(I) = (1/\langle I \rangle) \exp{(-I/ \langle I \rangle)}$, where $\langle I \rangle$ is the mean intensity.\\
Let us imagine to generate the pseudo-thermal field described above and to select through a pin-hole a certain number $\mu_f$ of spatial modes in its far field. Now, let us suppose that this pseudo-thermal field impinges on a second independent rotating ground-glass disk. The field that propagates after the second disk is named ``speckled-speckle field'' and, like the speckle field, is composed itself of spatial modes.
If $\mu_s$ speckles of this field are selected by a second pin-hole, the light intensity is characterized by the super-thermal statistics \cite{goodman}
\begin{equation}\label{densityfunctionsuper}
p_{sth}(I) = \frac{2(\mu_f \mu_s)^{\frac{(\mu_f + \mu_s)}{2}}}{\langle I \rangle \Gamma(\mu_f) \Gamma(\mu_s)} \left( \frac{I}{\langle I \rangle} \right)^{\frac{(\mu_f+\mu_s-2)}{2}} K_{|\mu_f-\mu_s| }\left( 2\sqrt{\mu_f \mu_s \frac{I}{\langle I \rangle} } \right),
\end{equation}
where $\langle I \rangle$ is the mean intensity of the speckled-speckle field, $\Gamma$ is the Gamma function, and $K_{|\mu_f-\mu_s|}$ is the $|\mu_f-\mu_s|$-order modified Bessel function of the second kind.\\
The $q$-th moments of this distribution are found to be 
\begin{equation}
\langle I^q \rangle = \left( \frac{\langle I \rangle}{\mu_f \mu_s} \right)^q \frac{\Gamma(\mu_f + q)\Gamma(\mu_s + q)}{\Gamma(\mu_f)\Gamma(\mu_s)}, 
\end{equation}
and in particular the second-order moment is
\begin{equation}
\langle I^2 \rangle = \langle I \rangle^2 \frac{(\mu_f + 1)(\mu_s + 1)}{\mu_f \mu_s}.
\end{equation}
This is useful to calculate the second-order autocorrelation function for intensity, that is
\begin{equation}\label{autocorr}
g^2(I) = \frac{\langle I^2 \rangle}{\langle I \rangle^2}  = \frac{(\mu_f + 1)(\mu_s +1)}{\mu_f \mu_s} = \left(1+ \frac{1}{\mu_f}\right)\left(1+ \frac{1}{\mu_s} \right).
\end{equation}
We notice that $g^2(I)$ is symmetric in the number of modes $\mu_f$ and $\mu_s$, and does not depend on the light intensity \cite{bianciardi}.
If $\mu_f = \mu_s = 1$, the autocorrelation function reaches the maximum value, that is $g^2(I) = 4$. Otherwise, when $\mu_f = \mu_s \rightarrow \infty$, $g^2(I) \rightarrow 1$.\\
In general, for the implementation of a GI protocol the calculation of the cross-correlation function is required since the optical scheme involves two replicas of the same field obtained by dividing it at a BS. 
\begin{figure}[h]
\centering
\includegraphics[width=10.5 cm]{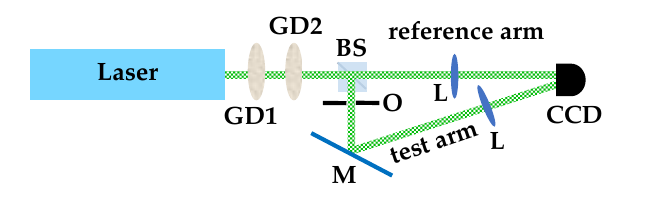}
\caption{Example of experimental scheme to realize a standard GI protocol with super-thermal light exploiting a spatial-resolving detector to detect both the test and the reference arm. GD1 and GD2: rotating ground-glass disks; BS: beam splitter; O: object; M: mirror; L: convergent lens; CCD: camera.
\label{GIscheme}}
\end{figure} 
As shown in Fig.~\ref{GIscheme}, one of the BS outputs, usually called test arm, interacts with the object O and is detected by a bucket detector, while the other output, usually called reference arm, is directly sent to a spatial-resolving detector without interacting with the object. Sometimes, for simplicity, the same spatial-resolving detector (e.g. a charged-coupled device (CCD) camera) can be used to detect the light coming from both arms. In this case, the effect of a bucket detector is obtained in post processing by summing all the pixels illuminated by the light coming from the test arm. The GI image is obtained by correlating the values of the bucket with the value of each pixel of the CCD camera over many repetitions. Assuming that a single pixel records no more than a single mode, namely that the speckle size is larger than that of a single pixel, the correlation function $G(I_i)$ can be written as
\begin{eqnarray}\label{Corrmatrix}
G (I_i) &=& \frac{\langle (\sum_{j = 1}^{\mu_s} I_j) I_i \rangle}{\langle \sum_{j = 1}^{\mu_s} I_j\rangle \langle I_i \rangle} = \frac{\langle I_i^2 \rangle + \langle \sum_{i \neq j}I_j I_i \rangle}{\langle \sum_{j = 1}^{\mu_s} I_j\rangle \langle I_i \rangle} = \nonumber\\
&=& \frac{\langle I^2 \rangle}{\langle I \rangle^2} \frac{\langle I_i \rangle}{\langle \sum_{j = i = 1}^{\mu_s} I_j\rangle} + \sum_{j \neq i =1}^{\mu_s} \frac{\langle I_i I_j \rangle}{\langle I_i \rangle \langle I_j \rangle} \frac{\langle I_j \rangle}{\langle \sum_{j = 1}^{\mu_s} I_j\rangle} =\nonumber\\
&=&  g^2(I) \frac{\langle I_i \rangle}{\langle \sum_{i = 1}^{\mu_s} I_i \rangle} + \sum_{j \neq i =1}^{\mu_s} g^{1,1}(I_i, I_j) \frac{\langle I_j \rangle}{\langle \sum_{j = 1}^{\mu_s} I_j \rangle},
\end{eqnarray}
where $g^{1,1}(I_i, I_j)$ is the cross-correlation function. Note that in Eq.~(\ref{Corrmatrix}) the quantity $I_i$ is the shot-by-shot intensity of a single mode, while $\langle \sum_{i = 1}^{\mu_s} I_i \rangle$ is the mean total intensity measured by the bucket detector, so that, assuming $\mu_s$-equally populated modes, we have that ${\langle I_i \rangle}/{\langle \sum_{i = 1}^{\mu_s} I_i \rangle} = 1/\mu_s$. According to Eq.~(\ref{autocorr}), $g^2(I) = (1 + 1/\mu_f)(1 + 1/\mu_s) = 2 (1 + 1/\mu_f)$, where we assumed $\mu_s = 1$ since this term corresponds to the case in which the pixel is correlated with itself. On the other hand, $g^{1,1}(I_i, I_j)$ gives the correlations between pixels that are different from each other, whose intensity is distributed according to that from the first disk, that is a multi-mode thermal distribution with $\mu_f$ modes. This means that
\begin{equation}\label{crosscorr} 
\sum_{j \neq i =1}^{\mu_s} g^{1,1}(I_i, I_j) = \sum_{j \neq i = 1}^{\mu_s} \frac{\langle I_i I_j \rangle}{\langle I_i \rangle \langle I_j \rangle} = (\mu_s -1) \left(1+ \frac{1}{\mu_f} \right) \frac{\langle I_i \rangle \langle I_j \rangle}{\langle I_i \rangle \langle I_j \rangle} = (\mu_s -1) \left(1+ \frac{1}{\mu_f} \right).
\end{equation}
Note that the value of $g^{1,1}(I_i, I_j) = (1+1/\mu_f)$ represents the background of the correlation image. The maximum value of this function, that is 2, is attained when $\mu_f = 1$, while the minimum value, that is 1, is achieved when $\mu_f \rightarrow \infty$.
By substituting Eq.~(\ref{crosscorr}) into Eq.~(\ref{Corrmatrix}), we obtain
\begin{eqnarray}\label{Corrmatrixfin}
G(I) &=& 2 \left( 1 + \frac{1}{\mu_f} \right) \frac{1}{\mu_s} + \frac{\mu_s -1}{\mu_s} \left( 1 + \frac{1}{\mu_f} \right)  = \left(1 + \frac{1}{\mu_f}\right) \left(1 + \frac{1}{\mu_s}\right),
\end{eqnarray}
that gives the general expression of the correlation function. Assuming $\mu_f = 1$, the two limit values are $G_{\rm max}(I) = 4$ for $\mu_s = 1$ and $G_{\rm min}(I) = 2$ for $\mu_s \rightarrow \infty$. For a direct comparison, we note that the expression of $G(I)$ in the case of thermal light is equal to $G(I) = (1 + 1/\mu)$, where $\mu$ is the number of detected speckles.\\ 
The calculation of $G$ is repeated for all pixels having coordinates ${k,l}$ of the CCD camera, so that a correlation matrix can be obtained:
\begin{equation} \label{GIcorrmatrix}
G_{\rm GI}(I_{k,l}) = \frac{\langle (\sum_{i,j = 1}^{M,N} I_{i,j}) I_{k,l} \rangle}{\langle \sum_{i,j = 1}^{M,N} I_{i,j} \rangle \langle I_{k,l} \rangle}, 
\end{equation}
under the assumption that the pixels corresponding to the bucket detector, having an area with size M$\times$N, are summed together.
As extensively discussed in the literature, to quantify the quality of the GI image, we need to consider some figures of merit. The most used are the visibility (V), the contrast (C) and the signal-to-noise ratio (SNR). For binary objects, they can be expressed in terms of $G(I)$ as follows \cite{losero,ferri10,iskhakov,ragy}
\begin{eqnarray}
{\rm V} &=& \frac{G_{\rm IN}(I) - G_{\rm OUT}(I)}{G_{\rm IN}(I) + G_{\rm OUT}(I)}\\
{\rm C} &=& \sqrt{G_{\rm IN}(I) - G_{\rm OUT}(I)}\\
{\rm SNR} &=& \frac{G_{\rm IN}(I) - G_{\rm OUT}(I)}{\sigma[G_{\rm OUT}(I)]},\label{snr} 
\end{eqnarray}
where $G_{\rm IN}(I)$ and $G_{\rm OUT}(I)$ are the values of the correlation inside and outside the GI image, while $\sigma[G_{\rm OUT}(I)]$ is the standard deviation of the part of the image that does not contain information about the object. It has already been demonstrated that the SNR is the most useful criterion, as it takes into account light fluctuations, quantifying the contribution of noise to the light signal. For instance, in the case of thermal light the maximum values of $G_{\rm IN}(I) = 2$, while $G_{\rm OUT}(I) = 1$, so that V$_{\rm th} = 1/3$. The same result can be achieved in the case of super-thermal light since the maximum value of $G_{\rm IN}(I)$ is equal to 4, while $G_{\rm OUT}(I) = 2$ if the speckled-speckle field is generated by a single speckle $\mu_f = 1$ exiting the first disk and entering the second one. This means that the visibility does not allow one to appreciate the difference between the two light states. For what concerns the contrast, in the case of thermal light with $\mu = 1$, C$_{\rm th}= \sqrt{2-1} = 1$, while for super-thermal light with $\mu_f = \mu_s = 1$, C$_{\rm sth} = \sqrt{4-2} = \sqrt{2}$. This proves that using two diffusers instead of one improves the contrast of the GI image. The definition of the contrast is connected to the SNR one, which can be rewritten as SNR=C$^2/\sigma[G_{\rm OUT}(I)]$. Indeed, this can be given in terms of different quantities, as remarked in [\citen{losero}]. According to our definition in Eq.~(\ref{snr}), the signal-to-noise ratio is also connected to the variance of the distribution of the background, considered in the denominator. In the case of thermal light, it has been demonstrated \cite{brida11} that SNR$_{\rm th}$ is proportional to $1/\sqrt{\mu}$. For what concerns super-thermal light, in analogy with pseudo-thermal light, we may state that, since $G(I)$ is a function of the number of modes $\mu_f$ and $\mu_s$ in the same way as it is a function of $\mu$ for thermal light, SNR$_{\rm sth}$ is proportional to $1/\sqrt{\mu_s}$ for $\mu_f = 1$. Of course, a difference between the two cases is expected in the proportionality coefficient, which is connected to the statistical properties of the two light fields \cite{bianciardi}. As already shown in Ref.~[\citen{PLA24}], our results seem to prove this statement. These considerations lead us to the conclusion that the values of the figures of merit do not depend on the intensity level of light, but only on its statistical properties. The main requirement is simply that the photodetector used to reveal the light is sensitive enough, or, in other words, that the chosen light level is well above the noise level of the detector. Therefore, the experimental results shown in the following Sections have been obtained by making full use of the dynamic range of the employed camera.\\
As remarked in the Introduction, the GI setup can be also exploited to perform DGI, since the difference between the two techniques is in data processing. We evaluate the correlation matrix in the case of DGI, which consists in subtracting from the GI correlation matrix the GI correlation matrix of the non-correlated part of the image. Thus, Eq.~(\ref{GIcorrmatrix}) is modified as follows
\begin{equation} \label{DGIcorrmatrix}
G_{\rm DGI}(I_{k,l}) = \frac{\langle (\sum_{i,j = 1}^{M,N} I_{i,j}) I_{k,l} \rangle}{\langle (\sum_{i,j = 1}^{M,N} I_{i,j})\rangle \langle I_{k,l} \rangle} - \frac{\langle (\sum_{p,q = 1}^{R,S} I_{p,q}) I_{k,l} \rangle}{\langle (\sum_{p,q = 1}^{R,S} I_{p,q})\rangle \langle I_{k,l} \rangle},
\end{equation}
where $\sum_{p,q = 1}^{R,S} I_{p,q}$ is the the sum of the intensities detected in a portion corresponding to the reference arm.\\
Concerning the figures of merits, we notice that, as discussed in Ref.~[\citen{ferri10}], the use of DGI instead of conventional GI can give better results in terms of SNR in the case of weakly absorbing objects. In the next Sections, we will show that for super-thermal light this advantage is preserved also in the non-absorbing case and that super-thermal light can give better results since its intensity fluctuations are larger with respect to those of thermal light.  
\subsection*{Characterization of the generated speckled-speckle field}
\begin{figure}[h]
\centering
\includegraphics[width=10.5 cm]{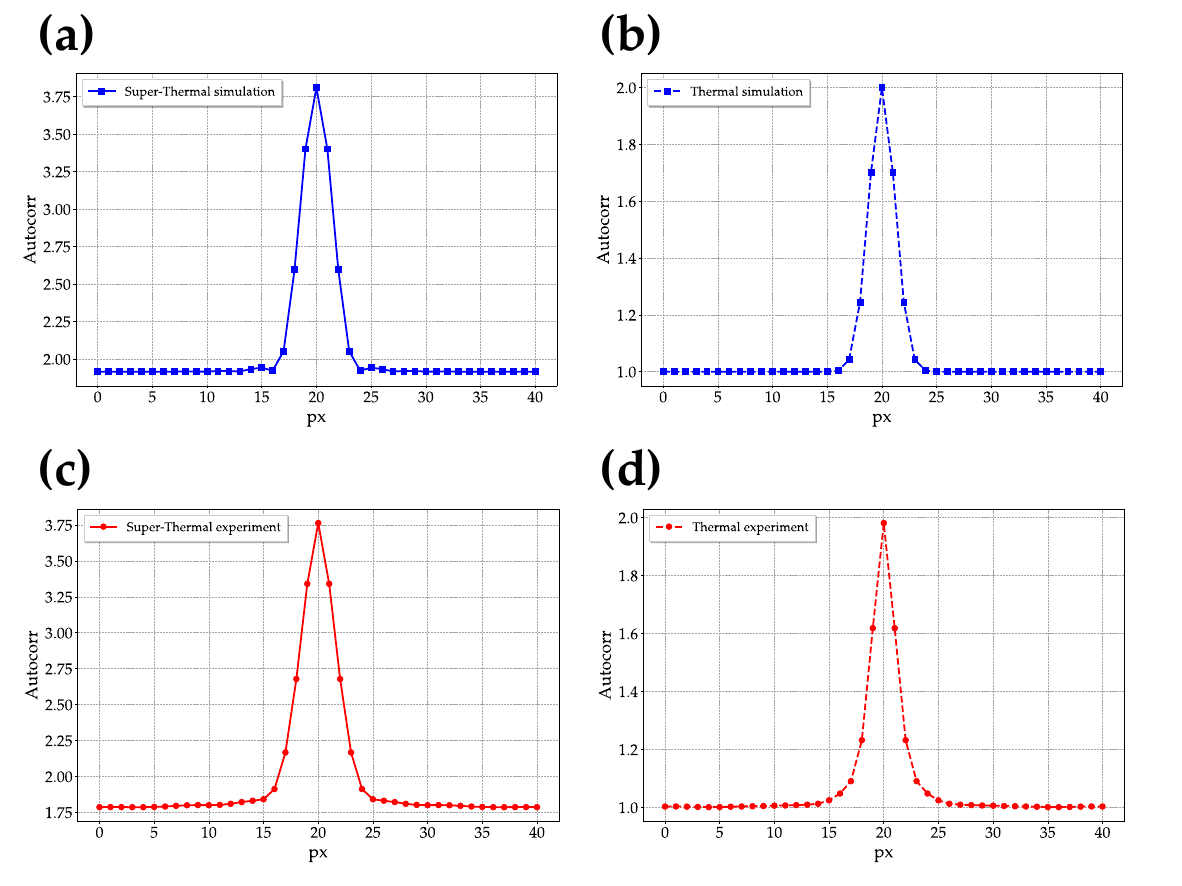}
\caption{Upper panels: section of the spatial autocorrelation function from simulated data in the case of super-thermal light (panel (a)) and pseudo-thermal light (panel (b)). Lower panels: the same as in the upper panels obtained from experimental data by considering the bucket side.  
\label{grautocorr}}
\end{figure}

\begin{figure}[hbtp]
\centering
\includegraphics[width=10.5 cm]{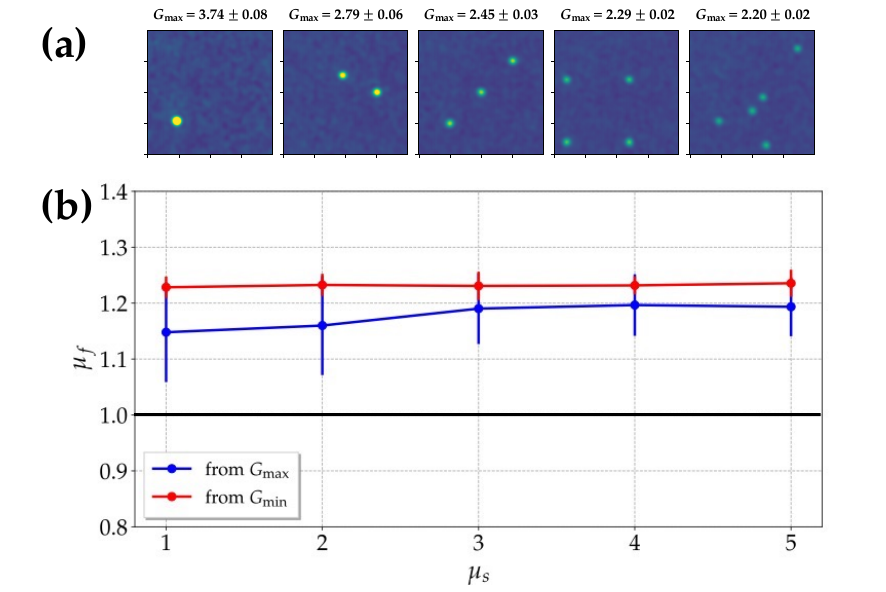}
\caption{Super-thermal light. Panel (a): autocorrelation images obtained by correlating 1, 2, 3, 4, and 5 pixels with all the pixels corresponding to the reference arm of the experimental images. The values of $G_{\rm max}$ indicated above the images are obtained by averaging over the corresponding number of pixels. Panel (b): $\mu_f$ as a function of the number of modes $\mu_s$ for super-thermal light. Blue dots + line: $\mu_f$ extracted from the maximum of $G(I)$; red dots + line: $\mu_f$ extracted from the minimum, $i.e.$ the background, of $G(I)$; black line: expected value of $\mu_f$ in the case of $G_{\rm max}(I)=4$ and $G_{\rm min}(I)=2$.    
\label{grcorrtemp}}
\end{figure}

\begin{figure}[hbtp]
\centering
\includegraphics[width=9.5 cm]{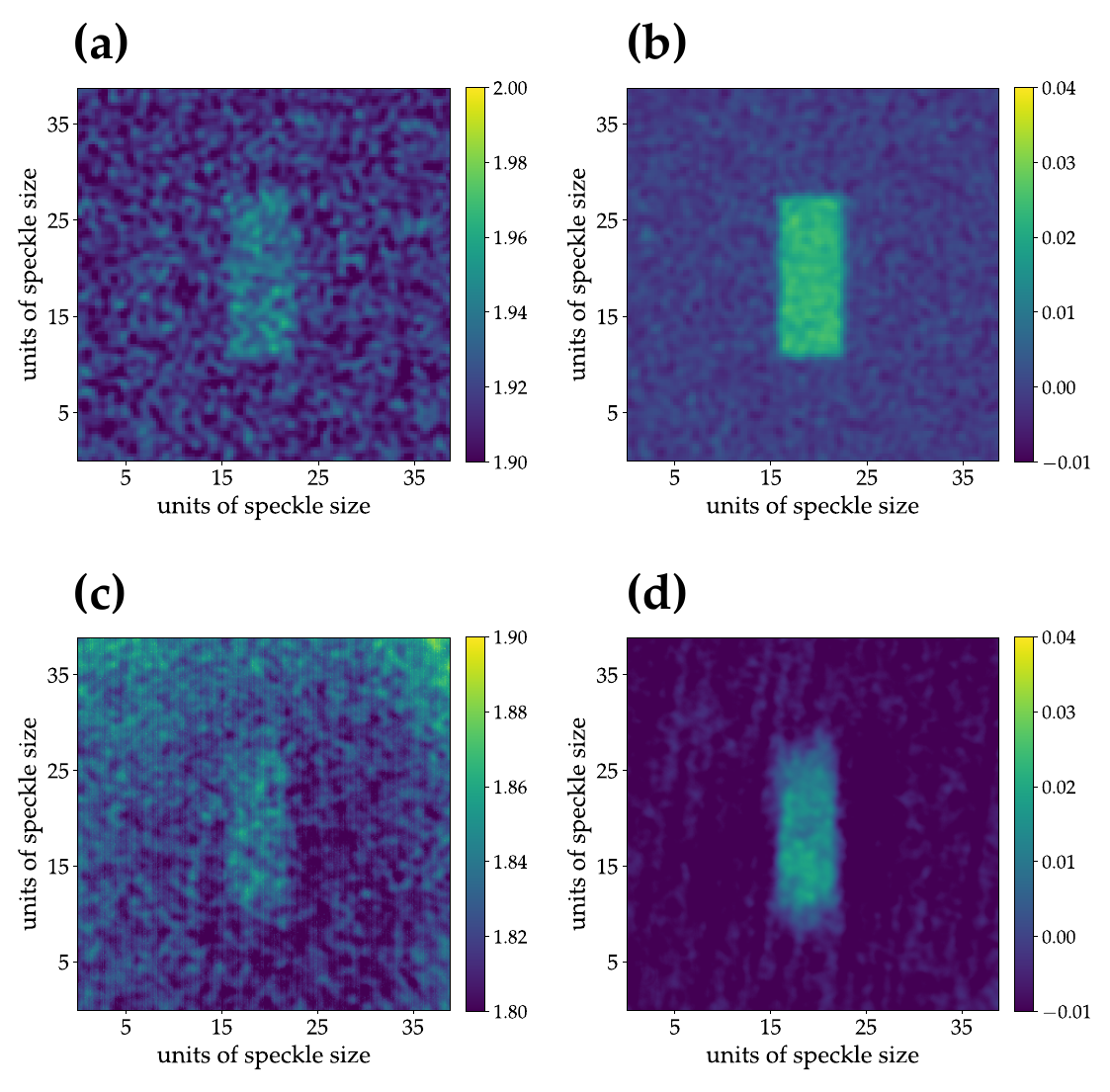}
\caption{Super-thermal light. Upper panels: GI (panel (a)) and DGI (panel (b)) images obtained by selecting an object 20 $\times$ 50 pixels large on the bucket detector from simulated data. Lower panels: the same as in upper panels from experimental data. The values of C are 0.150 in panel (a), 0.151 in panel (b), $0.13 \pm 0.03$ in panel (c), and $0.148 \pm 0.004$ in panel (d), while those of SNR are 2.74 in panel (a), 15.34 in panel (b), $2.3 \pm 0.9$ in panel (c), and $14.3 \pm 2.4$ in panel (d).  
 \label{imagesSTH}}
\end{figure}

\begin{figure}[hbtp]
\centering
\includegraphics[width=9.5 cm]{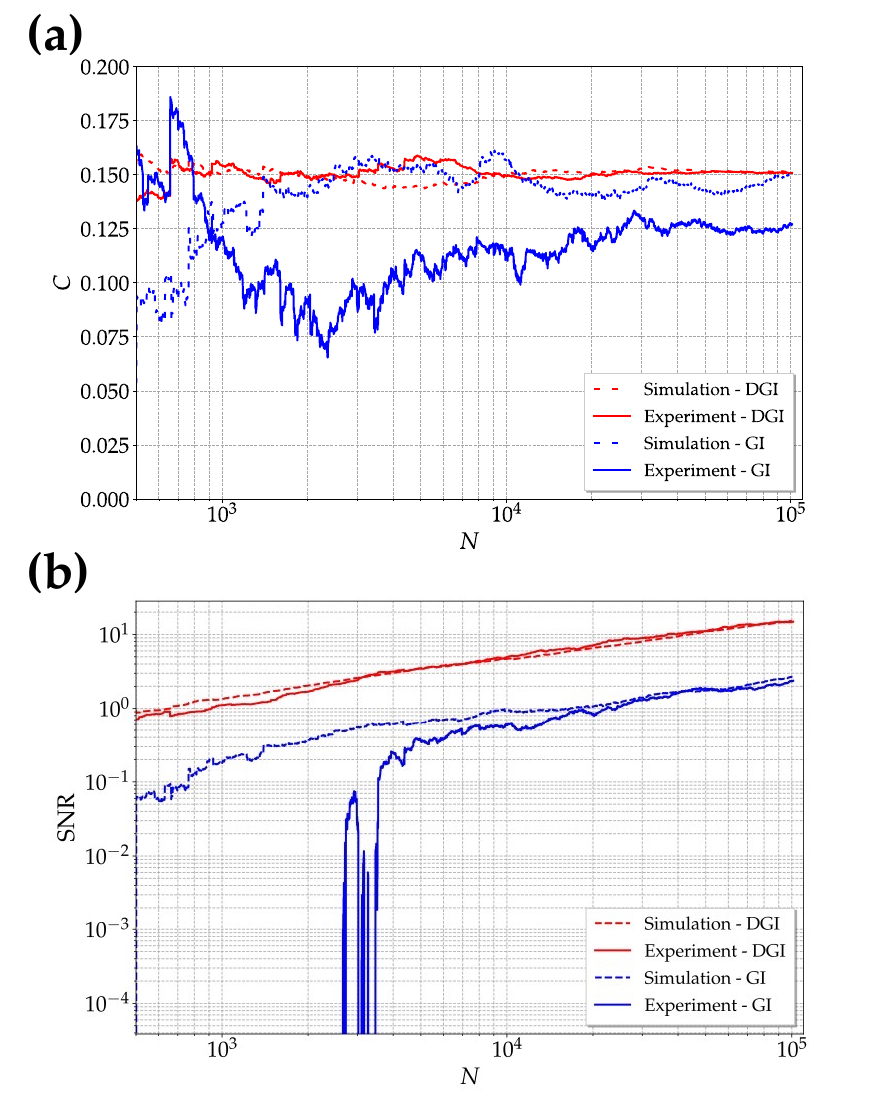}
\caption{Super-thermal light. C (panel (a)) and SNR (panel (b)) as functions of the number of images. Dotted lines: results from simulation; solid lines: results from experimental data. Red color refers to DGI, while blue color to GI. 
\label{grCSNRvsN}}
\end{figure}

To characterize the speckled-speckle field produced by both the simulation and the experiment, we firstly calculated the spatial autocorrelation function on each image and then averaged over the total number of images. In Fig.~\ref{grautocorr}, we show the section of the averaged autocorrelation image from simulations in the case of super-thermal light (panel (a)) and of pseudo-thermal light (panel (b)), and the analogous horizontal sections of the autocorrelation function obtained from the bucket portion of the experimental images (panels (c) and (d)).\\
From the spatial autocorrelation function some relevant information can be extracted, such as the type of employed light, the typical speckle size, and the number of modes selected by the first pin-hole in the case of super-thermal light. According to Eq.~(\ref{Corrmatrixfin}),
the maximum value of the autocorrelation function is equal to 4 and it is attained for $\mu_f = \mu_s = 1$, while the minimum, related to the background, is equal to 2 and is reached for $\mu_f = 1$ and $\mu_s \rightarrow \infty$. It is also worth noting that values smaller than 2 can be reached if $\mu_f >1$. In that case, also $G(I)$ will attain values smaller than 4. For what concerns pseudo-thermal light, the maximum value of autocorrelation function is 2, corresponding to the case $\mu = 1$, while the background is 1.\\
According to Fig.~\ref{grautocorr}(a) and (c), for super-thermal light, the peak of the autocorrelation function is equal to $3.81 \pm 0.03$ in the case of simulation and $3.76 \pm 0.04$ in the case of experiment, while the background is equal to $1.92 \pm 0.01$ and $1.79 \pm 0.02$, respectively. 
The direct comparison between simulation and experiment proves that the obtained results are compatible with each other, and in good agreement with theory. In particular, we note that in both panels ((a) and (c)) the value of the peak is slightly smaller than 4, thus corresponding to a number of modes $\mu_f$ slightly larger than 1. In particular, we got $\mu_f =$ 1.10 $\pm$ 0.02 and $1.14 \pm 0.03$ from the peak of the autocorrelation function referred to simulation and experiment, respectively, while we obtained $\mu_f =$ 1.09 $\pm$ 0.01 and $1.27 \pm 0.03$ from the values of the background. 
Note that these values are compatible with the analysis of the data in Figs. \ref{simul} and \ref{seqexp} in Section Methods.
Moreover, from the full width at half maximum of the function we can extract the typical size of the speckles of the speckled-speckle field, that is $d_{\rm ssp} = 4 \pm 1$ pixels for both simulation and the bucket side of the experimental images. If we repeat the same procedure on the reference arm of the experimental images, we get a width of the spatial autocorrelation function equal to $11 \pm 1$ pixels, in agreement with the different magnification existing between the two arms. The good correspondence between the values facilitates further comparisons in imaging applications, as it will become more evident in the next Section.\\
For what concerns pseudo-thermal light, according to Fig.~\ref{grautocorr}(b) and (d), the peaks of the autocorrelation functions are equal to $1.99998 \pm 0.00001$ and $2.08 \pm 0.02$, respectively, while the background is equal to 1 in both cases, corresponding to a single-mode thermal state. Moreover, from the width of the function, we can extract the typical size of the speckles of the speckle field, that is $d_{\rm sp} = 3 \pm 1$ pixels in both cases.\\
The number of modes $\mu_f$ selected by the first pin-hole can be also investigated by calculating the temporal autocorrelation function at different values of $\mu_s$ obtained by selecting a given number of pixels in the portion of the CCD camera illuminated by the light coming from the reference arm. First of all, we built a bucket detector by choosing 1 or more pixels at relative distance larger than the typical width of a speckle, and summed their intensities. Then, for each image, we correlated this sum with each pixel in the reference arm. In the resulting autocorrelation image we can recognize the presence of 1 or more speckles depending on the number of distinct selected pixels, as shown in Fig.~\ref{grcorrtemp}(a). 
The maxima of the autocorrelation image are found at the coordinates of the chosen pixels. The value of $\mu_f$ can be evaluated from Eq.~(\ref{Corrmatrixfin}) by setting $\mu_s$ equal to the number of selected pixels. In Fig.~\ref{grcorrtemp}(b) we show the value of $\mu_f$ as a function of the number of modes $\mu_s$ for super-thermal light. For a direct comparison, in the same Figure we also show the value of $\mu_f$ extracted from the value of the background, that is inverting $g^{1,1}(I_i, I_j) = (1+1/\mu_f)$. As observed in the case of the spatial autocorrelation function, we can notice that the two methods are not completely equivalent, at least for small values of $\mu_s$, even if they are compatible within 1$\sigma$. This discrepancy can be ascribed to the fact that the calculation of the number of modes $\mu_f$ from the maximum was obtained by repeating the procedure for a limited number of choices, by randomly choosing the pixels inside the portion of CCD camera. Conversely, the calculation of $\mu_f$ from the background was based on an average over several pixels. The difference between the two procedures is less evident in the case of large values of $\mu_s$ since the effect of possible fluctuations in the choice of the pixels to be correlated cancels out. Note that the obtained values of $\mu_f$ are slightly larger than 1, that is the expected value for $G_{\rm max}(I) = 4$ and $G_{\rm min}(I) = 2$. Nevertheless, as we will show in the following, this small discrepancy from the ideal case does not prevent us from investigating the possible advantages of super-thermal light. In more detail, we investigate the quality of the GI and DGI images as a function of the number of speckles illuminating the object, and as a function of the number of images, to understand what is the minimum number required to saturate the figures of merit.\\
In Fig.~\ref{imagesSTH} we show the GI and DGI images obtained by selecting an object 20 pixels $\times$ 50 pixels large on the bucket detector both in the case of simulation (panels (a) and (b)) and experiment (panels (c) and (d)).
We can notice that there is a good agreement between the results obtained from simulation and experiment for both strategies. This result can be ascribed to the fact that the values of $\mu_f$ and $\mu_s$ are compatible. 
Moreover, DGI images are sharper than the ones corresponding to GI, especially in the case of experimental data. 
\begin{figure}[hbtp]
\centering
\includegraphics[width=9.5 cm]{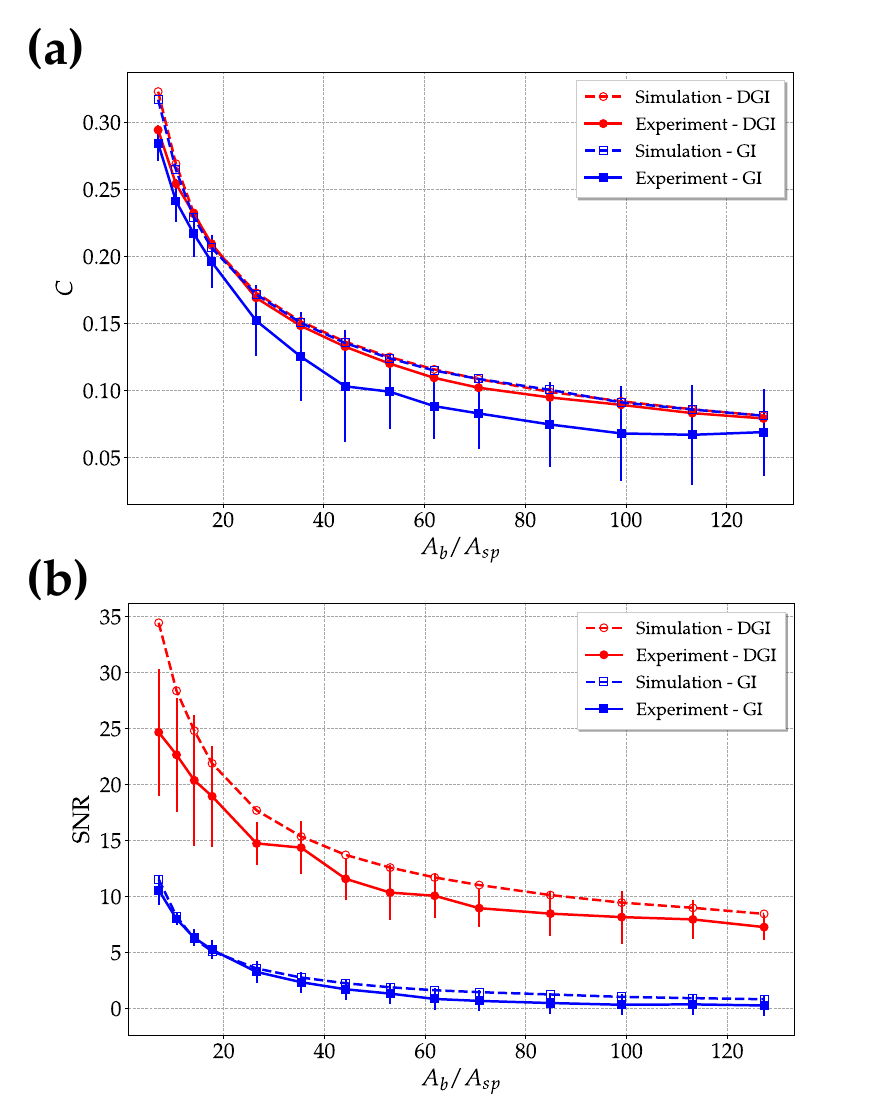}
\caption{Super-thermal light. C (panel (a)) and SNR (panel (b)) as functions of the ratio between the area of the object, $A_b$, selected on the bucket side and that of a typical speckle, $A_{sp}$. Open symbols + dotted lines: results from simulation; full symbols + solid lines: results from experimental data. Red color refers to DGI, while blue color to GI. The error bars corresponding to the experimental results were calculated by considering different areas of the correlation images for the evaluation of the background.
 \label{grCSNRsth}}
\end{figure} 
As anticipated in the Introduction, this is due to the fact that DGI technique removes the effect of noise due to the non-correlated part of the images. This can be quantified by the figures of merit. For the images in Fig.~\ref{imagesSTH} we obtain a SNR higher for DGI images than for GI ones, as explicitly indicated in the caption, for both simulated images and experimental ones. On the other hand, let us remark that C is independent of the chosen technique. This fact can be better appreciated by considering the values of C and SNR shown in Fig.~\ref{grCSNRsth} as a function of the ratio between the area of the object selected on the bucket side, $A_b$, and that of a typical speckle, $A_{sp}$, roughly corresponding to the number of modes $\mu_s$ illuminating the object. From the plot, we can notice that there is a larger increase of SNR with respect to C at different sizes of the object.
As a final investigation, we consider the behavior of C and SNR as a function of the number of images for a fixed choice of the object size, that is 20 $\times$ 50 pixels. We can clearly see from Fig.~\ref{grCSNRvsN} that, while the contrast attains a constant value with more than 10$^3$ images in the case of both GI and DGI, the SNR is still an increasing function even with 100,000 images, thus not reaching a saturation value.
Residual background fluctuations are evident in the case of GI technique: a number of images larger than 10$^4$ should be required to obtain a regular behavior, especially for the experimental realization, which is definitely more sensitive to nonidealities and optical distortions.
Note that $N = 100,000$ data are sufficient to reach good values of SNR ($>10$) in the case of DGI. \\ 
\section*{Discussion}
\begin{figure}[hbtp]
\centering
\includegraphics[width=14 cm]{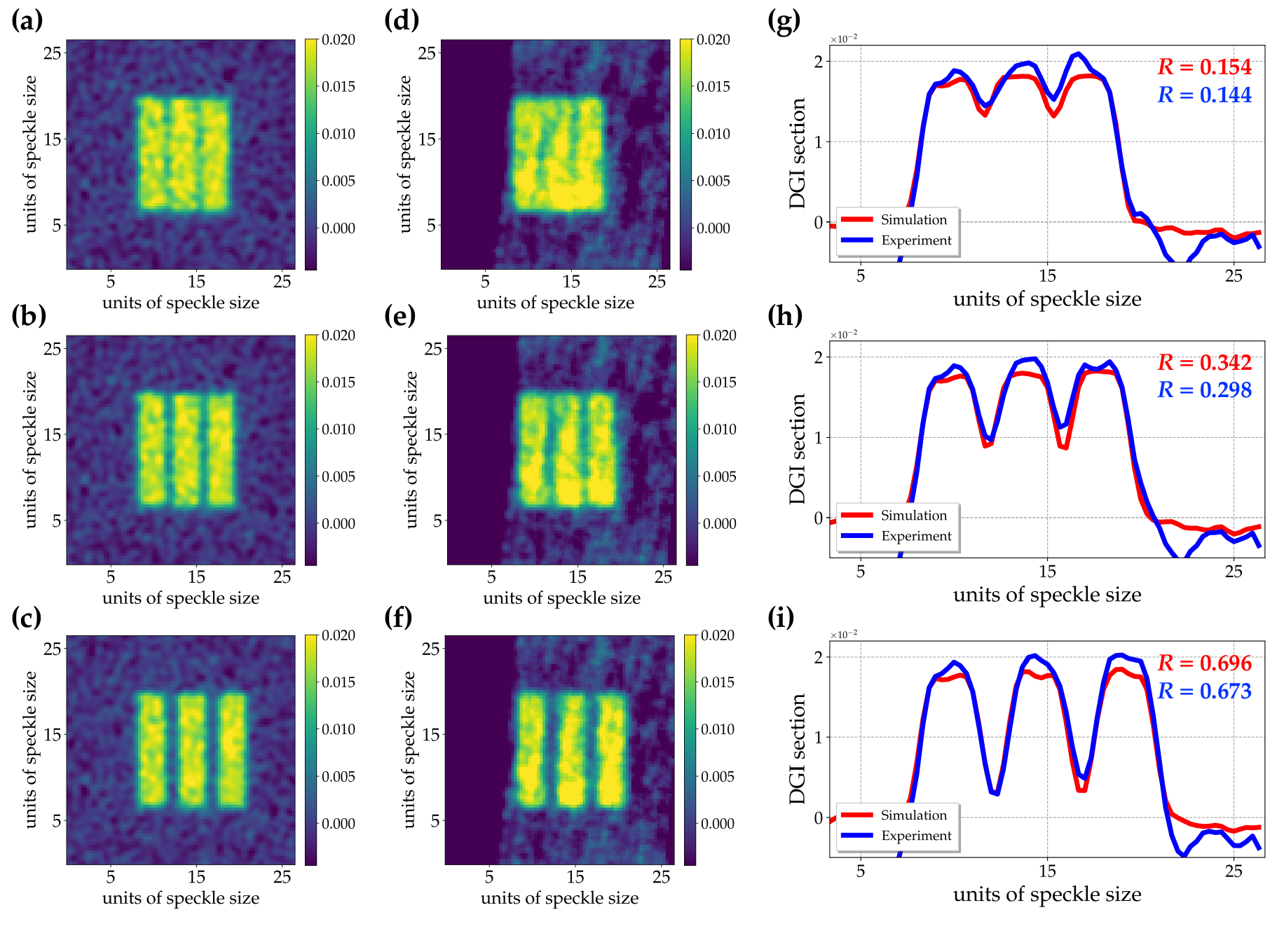}
\caption{Super-thermal light. (a) - (c): DGI images of 3-slit objects having a different pixel separation (1 pixel, 2 pixels, and 4 pixels, respectively) obtained from simulations. (d) - (f): DGI images of the same objects obtained from the experimental data. (g) - (i): sections of the DGI images, in which red color refers to simulations, while blue color to experiment. 
\label{3slitSTH}}
\end{figure}

The results shown above suggest different considerations. 
First of all, we notice that the performed simulation and the experimental data lead to very similar outcomes, which are also in good agreement with the predictions of the developed theoretical model. In particular, we verified the dependence of the correlation functions on the number of modes $\mu_f$ and $\mu_s$. 
Concerning the applications to imaging, we proved that in the case of super-thermal light there is always an advantage in using DGI instead of GI in terms of SNR. 
This result is different from what was obtained with pseudo-thermal light, as discussed in Ref.~[\citen{ferri10}]. In that case, Ferri et al. proved that DGI is better than GI only in the case of faint and weakly absorbing objects, while our analysis demonstrates that for super-thermal light the values of SNR for DGI are always larger than those for GI regardless of the size of the object. This is not the case of contrast, whose value as a function of the size of the object does not depend on the employed technique.\\
Even though in Ref.~[\citen{PLA24}] we have demonstrated that the absolute values of SNR that can be obtained by illuminating the object with pseudo-thermal light instead of super-thermal one are larger, the data shown in Fig.~\ref{grCSNRsth} prove that values of SNR close to 10 can be reached for a considerable number ($\sim 100$) of speckles illuminating the object. This good result is also accompanied by very good values of contrast, which definitely surpass the corresponding ones achieved with pseudo-thermal light.\\ 
An additional comparison between super-thermal light with respect to pseudo-thermal one can be performed in terms of image resolution by considering objects more complex than the one shown in Fig.~\ref{imagesSTH}. To compare pseudo-thermal and super-thermal lights, we implemented the DGI protocol by using the same number of images, namely $10^5$, even though, as stated in the comment to Fig.~\ref{grCSNRvsN}, DGI with super-thermal light would improve increasing further the number of images.
As to the objects, we considered three-slit binary masks having a size of 10 pixels $\times$ 40 pixels and a separation that varies from 1 pixel to 4 pixels. In panels (a) - (f) of Fig.~\ref{3slitSTH} we present the DGI autocorrelation images of the three objects obtained with super-thermal light from the simulations (panels (a) - (c)) and experimental data (panels (d) - (f)), respectively. 
To better investigate the results, in panels (g) - (i) we show the sections of the three images: red curves refer to simulated results, while blue curves to experimental ones. 
\begin{figure}[hbtp]
\centering
\includegraphics[width=14 cm]{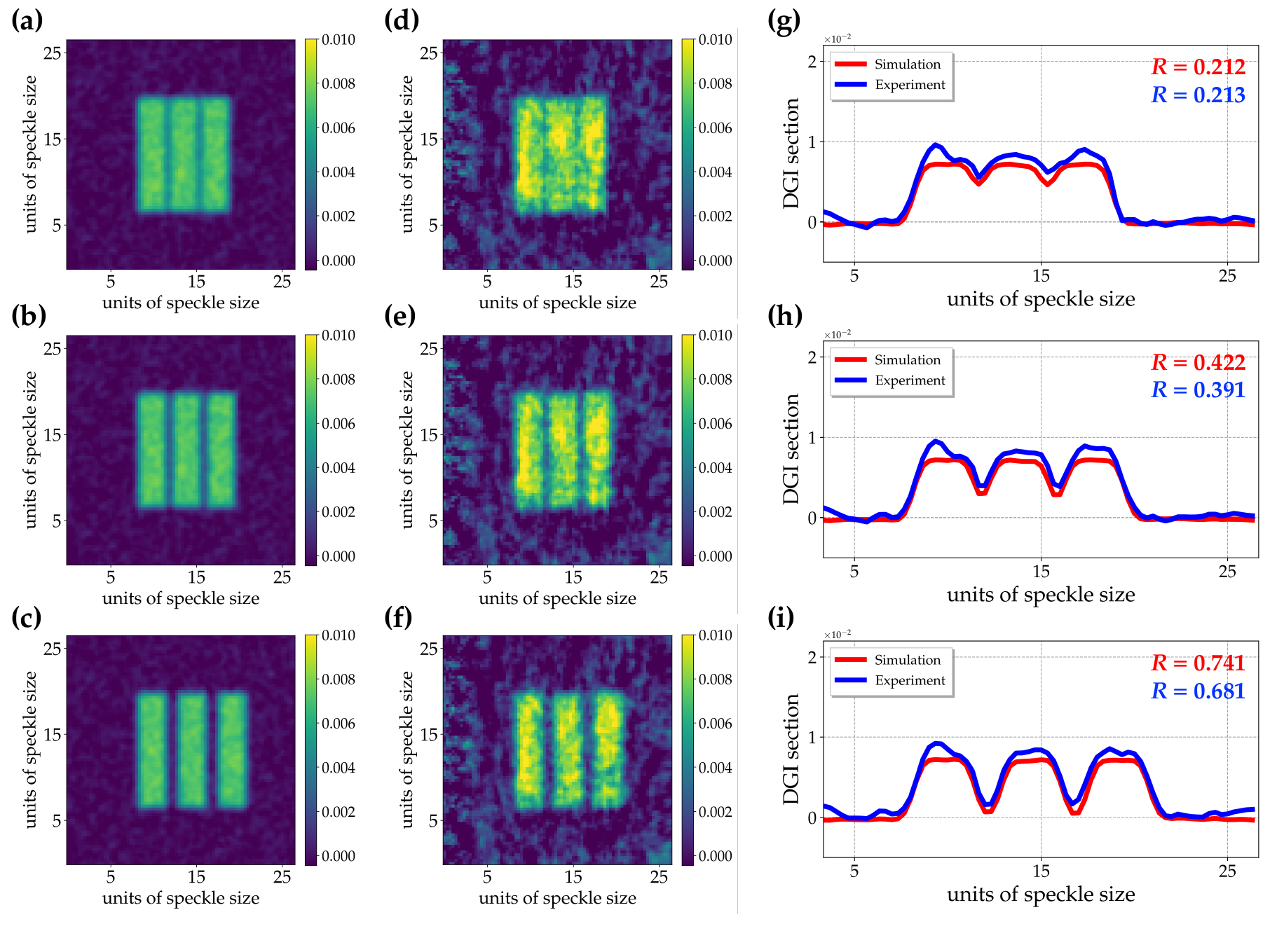}
\caption{Pseudo-thermal light. (a) - (c): DGI images of 3-slit objects having a different pixel separation (1 pixel, 2 pixels, and 4 pixels, respectively) obtained from simulations. (d) - (f): DGI images of the same objects obtained from the experimental data. (g) - (i): sections of the DGI images, in which red color refers to simulations, while blue color to experiment. 
\label{3slitTH}}
\end{figure} 
We can notice that the results shown in each panel are in good agreement. To quantify the image resolution, we define the resolution parameter as 
\begin{equation}
    R = (\langle y_{loc max} \rangle - \langle y_{loc min} \rangle)/ (\langle y_{loc max} \rangle + \langle y_{loc min} \rangle), \label{resolution}
\end{equation}
where $\langle y_{loc max} \rangle$ is the average of the local maxima of the DGI sections and $\langle y_{loc min} \rangle$ is the average of the neighbouring local minima of the DGI sections. In particular, we achieved the following results with super-thermal light: $R = 0.154$ from simulations and $R = 0.144$ from the experiment for 1-pixel separation, $R = 0.342$ from simulations and $R = 0.298$ from the experiment for 2-pixel separation, $R = 0.696$ from simulations and $R = 0.673$ from the experiment for 3-pixel separation. We ascribe the slightly higher values of $R$ in the case of simulated data to the total absence of non-idealities due to experimental misalignments. The analogous results obtained with pseudo-thermal light are shown in Fig.~\ref{3slitTH}: in panels (a) - (c) we present the DGI autocorrelation images from simulations, while in panels (d) - (f) those from experimental data. Again, in panels (g) - (i), we plot the sections of the images: red curves refer to simulated results, while blue curves to experimental ones. Even in this case we can notice a good agreement between simulation and experiment. Moreover, by simply looking at the maximum values of the sections, we can appreciate that higher values are obtained in the case of super-thermal light instead of pseudo-thermal one. Indeed, this behavior is connected to the higher values of $g^2$ and thus of C exhibited by super-thermal light. However, if we calculate the values of image resolution for the images in Fig.~\ref{3slitTH}, as done in the case of super-thermal light, we note that the obtained results are almost similar, thus proving that there is not a real advantage in using super-thermal light instead of pseudo-thermal one. In fact, with pseudo-thermal light we get: $R = 0.212$ from simulations and $R = 0.213$ from the experiment for 1-pixel separation, $R = 0.422$ from simulations and $R = 0.391$ from the experiment for 2-pixel separation, $R = 0.741$ from simulations and $R = 0.681$ from the experiment for 3-pixel separation.
This investigation leads us to conclude that, if the aim of a specific research is to detect the presence or absence of a target, super-thermal light, being it endowed with higher values of correlations, is definitely better than pseudo-thermal light. On the contrary, the distinguishability of fine details seems not to be improved with super-thermal light. 
As a matter of fact, these results are compatible with the nature of the super-thermal light studied in this work, which in terms of spatial correlation is in fact characterized by thermal statistics. Instead, in order to have an advantage in imaging applications, one should better exploit temporal correlation, which is not the case in GI, where the temporal component is simply mediated. Therefore, novel imaging protocols exploiting super-thermal statistics in the single-pixel should be conceived.
\section*{Conclusion}
In this work we investigated the usefulness of super-thermal light obtained by a laser beam passing through a sequence of two diffusers for imaging applications. We performed our analysis exploiting the model discussed in Ref.~[\citen{bianciardi}] and realizing both a numerical simulation and a real experiment. In particular, we proved that in both cases there is a good agreement with the theoretical expectations by investigating the role played by the numbers of modes selected at the exit of both the first rotating ground-glass disk and the second one.\\
We studied the quality of the reconstructed images in terms of contrast and SNR by employing both GI and DGI techniques. In general, both the simulation and the experimental realization proved that DGI offers many advantages with respect to GI, such as higher values of SNR and the requirement of a smaller number of images. In our work we also considered the image resolution, proving that the results are comparable to those achieved with pseudo-thermal light.\\
The performed analysis suggests further investigations of possible advantages offered by this kind of light. For instance, work is now in progress to check whether some benefits could be obtained by better exploiting the  temporal correlations exhibited by super-thermal light. 

\section*{Methods}
The super-thermal light described in this work has been produced in two different ways. First of all, we implemented a LabVIEW-based simulation of speckled-speckle fields, and calculated all the quantities introduced in Section Results. Secondly, we compared the obtained results with those achieved with an experimental realization of super-thermal light. In both cases, a direct comparison with pseudo-thermal light was performed.
\subsection{Numerical simulation}
The simulation was built with LabVIEW. The program generates a speckle field using $\delta-$correlated random matrices, which are then convolved with a Gaussian distribution to obtain a Gaussian field \cite{goodman}. 
\begin{figure}[hbtp]
\centering
\includegraphics[width=10.5 cm]{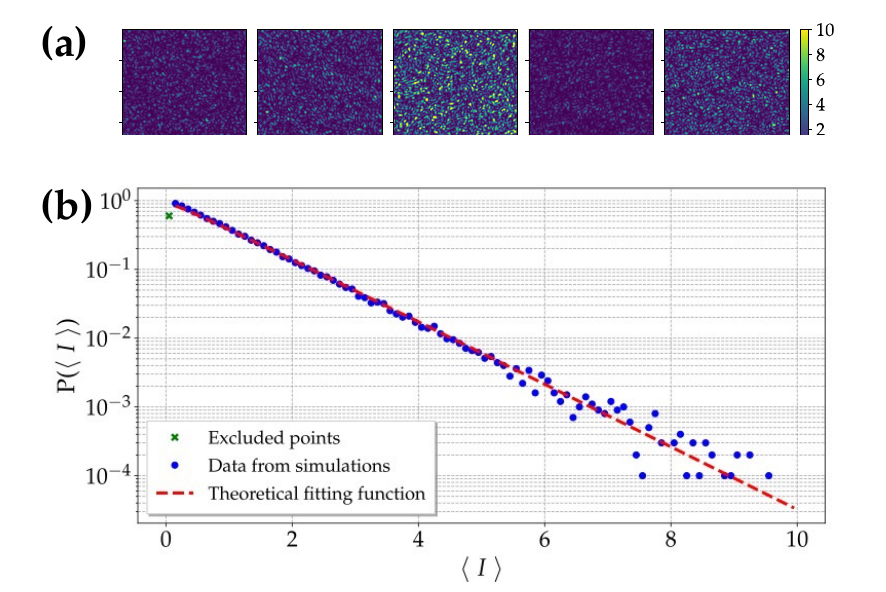}
\caption{Panel (a): simulated realizations of the speckled-speckle field at different mean intensities on a semi-logarithmic scale. Panel (b): probability distribution of the mean intensity of each image. Blue dots: experimental data; red dashed line: theoretical fitting function according to a  multi-mode thermal distribution, in which the number of modes $\mu_f$ is the only fitting parameter. The obtained value is equal to $\mu_f = 1.06 \pm 0.02$. One point has been excluded from the fit since its value is below the noise level.
\label{simul}}
\end{figure} 
The resulting matrices represent different realizations of the first speckle field. A portion of the field is then selected through a virtual pin-hole, PH, and used as the source of a second random process, that is the speckled-speckle field. To generate it, we used $N= 100$ scattering centers randomly moving in a three-dimensional space to scatter the light exiting the PH. The speckled-speckle field can be expressed as
\begin{equation}\label{speckled-specklefield}
E_2(x, y) = E_1(x, y) \exp{[i \phi{(r)}]} \exp{[i\phi_n(z)]},
\end{equation}
where $E_1(x, y) $ is the first speckle field, $\phi{(r)}$ is the phase of the speckled-speckle field, while the phase term $\phi_n(z)$ takes into account the thickness of the diffuser. 
The phase of the speckled-speckle field is related to the distance between each scattering center on the second disk and the position of a virtual spatial-resolving detector, such as a CCD camera:
\begin{equation}
\phi(r) = ikr =ik \sqrt{(x_{\rm CCD} - x_{\rm D2})^2 + (y_{\rm CCD} - y_{\rm D2})^2 + (z_{\rm CCD} - z_{\rm D2})^2},
\end{equation}
where $\{ x_{\rm D2},  y_{\rm D2},  z_{\rm D2} \}$ and $\{x_{\rm CCD}, y_{\rm CCD}, z_{\rm CCD} \}$ are the coordinates of the scattering center and CCD pixel, respectively.
The virtual CCD consists of $200 \times 200$ pixels. 
To replicate the real behavior of light, in the simulation we also consider the thickness of the second disk by inserting in Eq.~(\ref{speckled-specklefield}) the phase term 
\begin{equation}
\phi_n(z) = ik_n z_{\rm D2}, 
\end{equation}
in which $k_n = 2 \pi n/\lambda_0$ is the wavevector of light in the second diffuser with refractive index $n$, $\lambda_0$ is the wavelength of light in vacuum, while $z_{\rm D2}$ is the $z$-coordinate of each particle on the second diffuser, that is equal to the thickness of the disk. 
By using this strategy, we generated 100,000 realizations of the speckled-speckle field.
Some simulated realizations are shown in panel (a) of Fig.~\ref{simul}. By looking at the different intensities of the patterns we can appreciate the intensity fluctuations of the source, which is expected to be multi-mode thermal. This behavior can be quantified calculating the probability distribution of the mean intensity of each image. The resulting statistics is shown in panel (b) of Fig.~\ref{simul}, where the experimental data are presented together with the theoretical fitting function according to a  multi-mode thermal distribution, in which the number of modes $\mu_f$ is the only fitting parameter \cite{mandel}. The obtained value is equal to $\mu_f = 1.06 \pm 0.02$, which means that the field selected by the first pin-hole contains approximately a single mode, $i.e.$ a single speckle.
\subsection{Experimental implementation}
\begin{figure}[hbtp]
\centering
\includegraphics[width=10.5 cm]{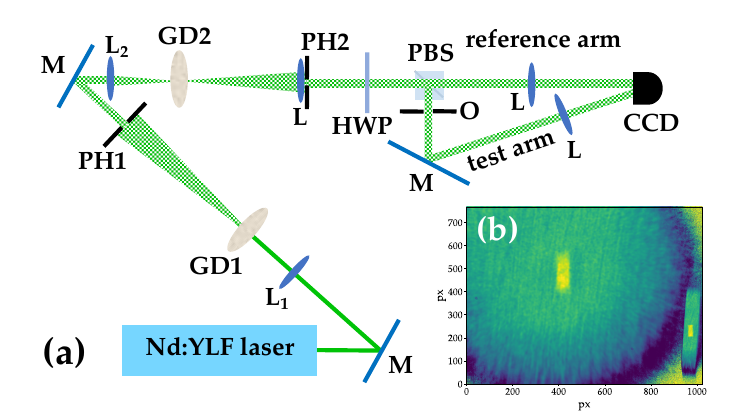}
\caption{Panel (a): sketch of the experimental setup. M: mirror; L$_1$: 200-mm-focal length lens; L$_2$: 100-mm-focal length lens; GD1 and GD2: rotating ground-glass disks; PH1: 2.5-mm-diameter pin-hole; PH2: 4-mm-diameter pin-hole; HWP: half-wave plate; PBS: polarizing cube beam splitter; L: 100-mm-focal length lens; O: object; CCD: camera. Panel (b): DGI image consisting of two parts: the autocorrelation image on the bucket side on the right and the cross-correlation one corresponding to the reference arm on the left. 
\label{setup}}
\end{figure}

\begin{figure}[hbtp]
\centering
\includegraphics[width=10.5 cm]{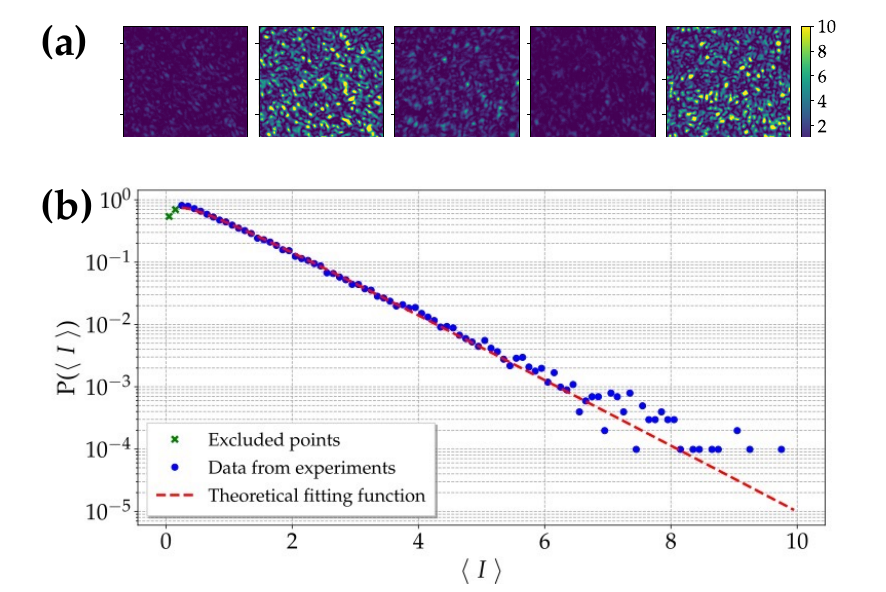}
\caption{Panel (a): experimental realizations of the speckled-speckle field at different mean intensities on a semi-logarithmic scale. Panel (b): probability distribution of the mean intensity of each image. Blue dots: experimental data; red dashed line: theoretical fitting function according to a  multi-mode thermal distribution, in which the number of modes $\mu_f$ is the only fitting parameter. The obtained value is equal to $\mu_f = 1.25 \pm 0.02$. Two points have been excluded from the fit since their value is within the noise level of the CCD camera.
\label{seqexp}}
\end{figure}

As sketched in Fig.~\ref{setup}(a), the second-harmonic pulses (at 523 nm, 5-ps pulse duration) of a Nd:YLF laser regeneratively amplified at 500 Hz were focused on  the surface of a rotating ground-glass disk, GD1. We selected a portion of the speckle field in far field by means of a pin-hole, PH1, having a diameter of 2.5 mm. This choice roughly corresponds to selecting a single speckle. The light passing through PH1 was then focused on the surface of a second rotating ground-glass disk, GD2.
The far-field condition of the speckled-speckle field was achieved by putting a lens with a 100-mm focal-length 100 mm behind GD2 so that the speckle field propagated with negligible divergence. The pattern was then split into two parts by a system composed of a half-waveplate (HWP) and a polarizing cube beam splitter (PBS) used to finely tune the balancing between the intensities in the two output arms. The reflected output was used as the test arm. An object consisting of a single slit with a 0.8-mm diameter was placed at 3.5 cm from the PBS and an imaging system with a magnification approximately equal to 1/3 was built using a 100-mm focal-length lens. The image was formed on a portion of a CCD camera (DCU223M, Thorlabs, 1024 $\times$  768 squared pixels, 4.65-$\mu$m pixel pitch). On the transmitted arm a 1:1 image of the speckle field at 3.5 cm from the PBS was built using another 100-mm focal-length lens. The image was formed on a different portion of the same CCD camera. 
Some typical single-shot images of the reference arm are shown in panel (a) of Fig.~\ref{seqexp}. 
Note that the speckles appear larger (roughly by a factor of 3) than those in the equivalent sequence of images from simulation shown in Fig.~\ref{simul}. This is due to the fact that the size of the simulated speckles was chosen similar to that of the experimental speckles of the bucket side, which were demagnified by a factor of approximately $M = 1/3$. In panel (b) we also present the probability distribution of the mean intensity of each experimental image together with the theoretical fitting function according to a  multi-mode thermal distribution, in which the number of modes $\mu_f$ is the only fitting parameter. The obtained value is equal to $\mu_f = 1.25 \pm 0.02$, which is slightly larger than the analogous value obtained from simulations. Nevertheless, as shown in Section Results, this small discrepancy does not prevent a direct comparison between simulated data and experimental ones.\\ 
To investigate the minimum number of images required to obtain either a GI image or a DGI one, and to evaluate their quality in terms of the already-mentioned figures of merit, we saved 100,000 realizations of the speckled-speckle field. As explained in Section Results, the calculation of the correlation matrix (see Eq.~(\ref{DGIcorrmatrix})) over this number of realizations contains the autocorrelation image on the bucket side (see the right side of Fig.~\ref{setup}(b)) and the cross-correlation image on the reference arm (see the left side of the same Figure).\\
For a direct comparison, we repeated the experiment with pseudo-thermal light by removing GD1 and adjusting the divergence of the beam impinging on GD2 in order to obtain a speckle field with speckles having roughly the same size as those obtained with speckled-speckle light. Also in this case, we saved 100,000 images.\\

Data availability
The datasets generated and analyzed during the current study are available from the corresponding authors on reasonable request.




\section*{Acknowledgements}
The authors thank Fabio Ferri, Alberto Parola and Camilla Bianciardi (University of Insubria) for fruitful discussions. S.C. and A.A. acknowledge the support by PNRR D.D.M.M. 351/2022. G.C. acknowledges the financial support of the INFN through the project QUANTUM. This work was financially supported by PNRR MUR Project PE0000023-NQSTI.

\section*{Author contributions statement}
S.C. and G.C. contributed equally to this work. A.A. and M.B. conceived the experiment; A.A. and S.C. conducted the experiment; G.C. performed the simulations; S.C. and G.C. analyzed the results. All authors reviewed the manuscript. 

\section*{Additional information}
\textbf{Competing interests}
The authors declare no conflict of interest.

\end{document}